\documentclass[prb,twocolumn,showpacs,superscriptaddress,amsmath]{revtex4}
\usepackage{graphicx}
\usepackage{amssymb}
\usepackage{citesort}

\begin{document}
\title{The Paired Electron Crystal: order from frustration in the
quarter-filled band}
\author{S. Dayal}
\affiliation{Department of Physics and Astronomy and HPC$^2$ Center for 
Computational Sciences, Mississippi State University, Mississippi State MS 39762}
\author{R.T. Clay}
\affiliation{Department of Physics and Astronomy and HPC$^2$ Center for 
Computational Sciences, Mississippi State University, Mississippi State MS 39762}
\author{H. Li}
\affiliation{Department of Physics, University of Arizona, Tucson, AZ 85721}
\author{S. Mazumdar}
\affiliation{Department of Physics, University of Arizona, Tucson, AZ 85721}
\date{\today}
\begin{abstract}
We present a study of the effects of simultaneous charge- and
spin-frustration on the two-dimensional strongly correlated
quarter-filled band on an anisotropic triangular lattice. The
broken-symmetry states that dominate in the weakly frustrated region
near the rectangular lattice limit are the well known
antiferromagnetic state with in-phase lattice dimerization along one
direction, and the Wigner crystal state with the checkerboard charge
order. For moderate to strong frustration, however, the dominant phase
is a novel spin-singlet paired-electron crystal (PEC), consisting of
pairs of charge-rich sites separated by pairs of charge-poor
sites. The PEC, with coexisting charge-order and spin-gap in two
dimension, is the quarter-filled band equivalent of the valence bond
solid (VBS) that can appear in the frustrated half-filled band within
antiferromagnetic spin Hamiltonians.  We discuss the phase diagram as
a function of on-site and intersite Coulomb interactions as well as
electron-phonon coupling strength. We speculate that the spin-bonded
pairs of the PEC can become mobile for even stronger frustration,
giving rise to a paired-electron liquid. We discuss the implications
of the PEC concept for understanding several classes of quarter-filled
band materials that display unconventional superconductivity, focusing
in particular on organic charge transfer solids.  Our work points out
the need to go beyond quantum spin liquid (QSL) concepts for highly
frustrated organic charge-transfer solids such as
$\kappa$-(BEDT-TTF)$_2$Cu$_2$(CN)$_3$ and
EtMe$_3$Sb[Pd(dmit)$_2$]$_2$, which we believe show
frustration-induced charge disproportionation at low temperatures.  We
discuss possible application to layered cobaltates and
$\frac{1}{4}$-filled band spinels.
\end{abstract}

\maketitle

\section{Introduction}
\label{intro}

Strong Coulomb electron-electron (e-e) interactions can drive
transitions from metallic to exotic insulating states, the most well
known of which are the Mott-Hubbard semiconductor (MHS) and the Wigner
crystal (WC). The MHS is a characteristic of systems with carrier
concentration per site $\rho=1$ and is driven by strong onsite e-e
repulsion, the Hubbard $U$ interaction. Depending upon the lattice
structure the critical $U$ at which the metal-insulator (MI)
transition occurs, $U_c$, can be 0$^+$ or finite
\cite{Lieb68a,Morita02a,Ohashi06a}.  In contrast, the WC occurs in
systems with $\rho\neq 1$, and is characterized by charge-ordering
(CO) , {\it i.e.}, a periodic arrangement of {\it single} charge
carriers on the lattice. The WC is driven by strong onsite as well as
inter-site Coulomb interactions \cite{Hubbard78a}. Although in
principle the WC is likely at any arbitrary $\rho$, it has been been
studied most intensively for $\rho=\frac{1}{2}$ bipartite lattices,
where the nearest neighbor (n.n) Coulomb repulsion can drive the MI
transition \cite{Penc94a,Seo97a}. The combined effects of e-e and
electron-phonon (e-p) interactions are also of interest, usually in
one dimension (1D), where the MHS can further exhibit the spin-Peierls
(SP) transition. Importantly for our purpose here the above
semiconducting states have been intensively studied over the past
several decades, and are largely understood, although arguments
regarding the magnitude of $U_c$ for formation of the MHS state in
specific lattices or the detailed mechanism of the MI transition may
continue to persist. In the present work, we discuss a new
correlated-electron semiconductor, the paired-electron crystal (PEC),
that occurs in $\rho=\frac{1}{2}$ systems in the presence of moderate
to strong geometric lattice frustration \cite{Li10a}. We believe that
our work has direct application to 2:1 cationic or 1:2 anionic
charge-transfer solids (CTS) that exhibit correlated insulator--SC
transitions, and further applies to other inorganic
strongly-correlated $\frac{1}{4}$-filled materials.

The combined effects of e-e interactions and geometric lattice
frustration are of strong current interest
\cite{Moessner06a,Balents10a}.  The bulk of the work here is for
$\rho=1$, where the Hubbard model in the limit $U\rightarrow\infty$
reduces to the Heisenberg spin Hamiltonian. Interest in the
consequences of lattice frustration stems from the seminal proposal by
Anderson that the ground state of the Heisenberg antiferromagnet (HAF)
model on a triangular lattice is a quantum spin liquid (QSL) with no
spin ordering even at zero temperature\cite{Anderson73a,Fazekas74a}.
The type of wavefunction usually assumed to describe a QSL is often
referred to as a resonating valence bond (RVB) state.  Whether or not
RVB states appear in the square lattice for $\rho$ slightly different
from 1, and the relationship of such states to superconductivity (SC)
in doped strongly correlated semiconductors remains contentious. An
extension of the RVB theory of dopant-induced SC in $\rho \neq 1$ is
the proposal that frustration-induced SC occurs in the anisotropic
triangular lattice within the simple Hubbard model even for $\rho$
exactly 1, where a narrow superconducting phase is straddled on both
sides by broader paramagnetic metallic (PM) and antiferromagnetic (AFM)
insulator  phases
\cite{Vojta99a,Schmalian98a,Kino98a,Kondo98a,Powell0507,Baskaran03a,Gan05a,Gan06a,Sahebsara06a,Watanabe06a,Kyung06a}.
It has been claimed that this transition explains the SC in the
CTS
\cite{Ishiguro,Vojta99a,Schmalian98a,Kino98a,Kondo98a,Powell0507,Baskaran03a,Gan05a,Gan06a,Sahebsara06a,Watanabe06a,Kyung06a}.
Recent numerical work by us and others, however, have
determined that SC is absent within the $\rho=1$ triangular lattice
Hubbard model \cite{Mizusaki06a,Clay08a,Tocchio09a} and the earlier
results are artifacts of mean-field approximations.

While the ground state of the HAF on the isotropic triangular lattice
is now known to be the ordered 120$^\circ$ AFM rather than the
originally proposed QSL state \cite{Huse88a,Jolicoeur89a,Singh92a},
other frustrated lattices, most notably the Kagom\'e lattice
\cite{Zeng90a}, have been investigated in the search for QSL states.
Proposed ground states here include various types of QSL states
\cite{Ryu07a,Ran07a,Hermele08a} as well as valence-bond solid (VBS) states
\cite{Hastings00a,Budnik04a,Singh,Evenbly10a}.  The literature on VBS
states has a long history going back to the well known Ghosh-Majumdar
model \cite{Majumdar69}.  The common theme in works on VBS is the
frustration-driven transition from the AFM state to a total spin $S=0$
singlet state. We have found a similar frustration-driven transition
from the AFM to a $S=0$ state in strongly correlated systems with
$\rho=\frac{1}{2}$, where reduction to a spin Hamiltonian is not
possible.

In contrast to the voluminous literature on correlated and frustrated
systems at $\rho=1$, the literature on frustrated $\rho \neq 1$ is
relatively sparse and new.  The discovery of SC\cite{Takada03b} in hydrated
Na$_x$CoO$_2$ has spurred interest in correlated
systems away from $\rho=1$ \cite{Motrunich04a,Choy07a,Merino09b}, although to the best of our
knowledge only isotropic triangular lattices have been studied. We
will specifically focus on $\rho=\frac{1}{2}$ within the present
work---on the triangular lattice with varying anisotropy. In the
square lattice limit at this density, we show that spontaneous
in-phase dimerization occurs in the presence of electron-phonon
interactions modulating n.n hopping integrals, leading to an {\it
  effective} $\rho=1$ system with one electron per dimer and AFM
order. This result is the origin of the so-called the dimer
Mott-Hubbard model \cite{Kino,McKenzie97a} that is commonly used to
describe the 2:1 cationic or 1:2 anionic organic CTS.  Very recently
we have proposed that under the influence of lattice frustration this
dimer Mott-Hubbard AFM state gives way to a spin-paired state that we
termed the PEC \cite{Li10a}. The PEC is
different from any of the above more well known correlated
semiconducting states in that it is a WC of Heitler-London
spin-singlets---simultaneously charged-ordered and spin
$S=0$. Alternately, the PEC is the $\rho=\frac{1}{2}$ equivalent of
the $\rho=1$ VBS. A conceptually similar state was postulated for the
electron gas many years back by Moulopoulos and Ashcroft
\cite{Moulopoulos92a,Moulopoulos93a}. There is a fundamental
similarity between this earlier work and ours, in that in both cases
the pairing is driven by the exchange interaction. Our previous work
\cite{Li10a} only considered a limited set of parameters, and the full
phase diagram was not discussed. Here we give a more complete phase
diagram, including the competition with the WC state that was ignored
before. As with the VBS state at $\rho=1$, the PEC is a consequence of
frustration-induced quantum effects. Its extraordinary stability at
$\rho=\frac{1}{2}$ is a commensurability effect (recall that MHS and
WC formation also require commensurability).

In section \ref{model} we introduce the model we consider. We include
e-p interactions to stabilize the lattice dimerization that gives the
effective $\rho=1$ dimer lattice. We present the physical mechanism
behind the PEC formation by briefly discussing simple molecular
clusters, for which we show that spin-singlet formation in
$\rho=\frac{1}{2}$ necessarily requires charge disproportionation. The
charge disproportionation in the infinite one-dimensional (1D) chain
and the so-called zigzag ladder leads to periodic CO, viz., the
simplest PECs. Following the discussions of these simple cases, we
introduce the two-dimensional (2D) lattice that will be the focus of
this work, and discuss the different possible phases. In section
\ref{results} we present out numerical results for the 2D system,
covering a wide region of parameter space.  In section
\ref{discussion} we discuss the relevance of our results for several
classes of $\rho=\frac{1}{2}$ materials and the outlook for
understanding unconventional SC. We particularly emphasize the cases
of the organic charge-transfer solids CTS
$\kappa$-(BEDT-TTF)$_2$Cu$_2$(CN)$_3$ and EtMe$_3$Sb[Pd(dmit)$_2$]$_2$
which have been described as QSLs within the effective $\rho=1$
scenario
\cite{Lee05a,Motrunich05a,Galitski07a,Lee07a,Lee08a,Qi08a,Qi09a,Xu09a,Grover10a}. We
believe frustration-induced charge disproportionation is an alternate
possibility.

\section{Theoretical model}
\label{model}

\subsection{Hamiltonian}
\label{afm-pec} 

The Hamiltonian we consider
contains electron hopping, semi-classical inter-site and
onsite e-p couplings, and onsite and n.n. Coulomb
interactions:
\begin{eqnarray}
H&=&-\sum_{\nu,\langle ij\rangle_\nu}t_\nu(1+\alpha_\nu\Delta_{ij})B_{ij} \label{ham} 
+\frac{1}{2}\sum_{\nu,\langle ij\rangle_\nu} K^\nu_\alpha \Delta_{ij}^2 \\
&+&\beta \sum_i v_i n_i + \frac{1}{2}K_\beta \sum_i v_i^2  \nonumber \\
&+& U\sum_i n_{i\uparrow}n_{i\downarrow} + 
\frac{1}{2}\sum_{\langle ij\rangle}V_{ij} n_i n_j. \nonumber 
\end{eqnarray}
In Eq.~\ref{ham}, $\nu$ indexes the different bond directions in the
lattice; for example $\nu=x$ in 1D and $\nu=\{x,y\}$ in the 2D square
lattice.  Our actual calculations (see below) are for the anisotropic
triangular lattice, $\nu=\{x,y,x+y\}$.
$B_{ij}=\sum_\sigma(c^\dagger_{i\sigma}c_{j\sigma}+ H.c.)$ is the
electron hopping between sites $i$ and $j$ with electron creation
(annihilation) operators $c^\dagger_{i\sigma}$ ($c_{i\sigma}$).
$\alpha_\nu$ is the inter-site e-p coupling constant, $K^\nu_\alpha$
is the corresponding spring constant, and $\Delta_{ij}$ is the
distortion of the bond between sites $i$ and $j$.  $v_i$ is the
intra-site phonon coordinate and $\beta$ is the intra-site e-p
coupling with corresponding spring constant $K_\beta$.  Both
$\Delta_{ij}$ and $v_i$ are determined self-consistently
\cite{Clay03a}.  $\alpha_\nu$ are in general taken close to the
minimum value needed for the transition to occur, our goal being the
replication of the same instability from finite cluster calculations
that would occur in the infinite system for $0^+$ coupling.  $U$ and
$V_{ij}$ are on-site and n.n. Coulomb interactions, respectively.  The
physically relevant range of $V_{ij}$ is $V_{ij}<\frac{U}{2}$ based on
comparison between $\rho=1$ and $\rho=\frac{1}{2}$ CTS \cite{Clay07a}.

\subsection{Coupled spin-singlet and CO at $\rho=\frac{1}{2}$}
\label{pec-motiv}

\begin{figure}[tb]
\centerline{\resizebox{3.3in}{!}{\includegraphics{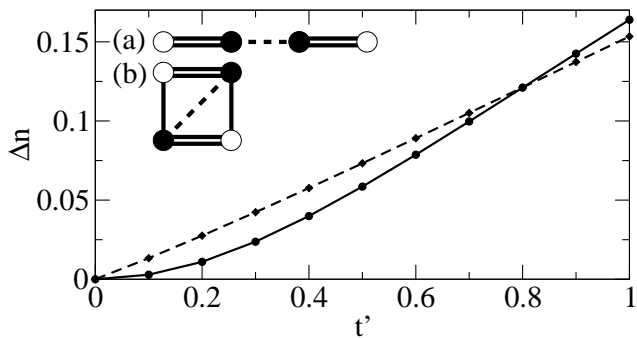}}}
\caption{Charge difference $\Delta n$ between members of the same
  dimer as a function of the hopping integral $t^\prime$ corresponding
  to the dotted bonds, for the two molecules (a) and (b) given as
  inserts. Each 4-atom molecule contains two electrons.
The intra-dimer double bonds have strength
  $t_1=1.5$, and the single bonds in (b) are $t_2=0.5$.  The
  results shown are for $U=4$ and $V=0$.  Solid (dashed) curves
 show $\Delta n$ for the linear (square) molecules. Lines are
  guides to the eye.  Filled and empty circles of the molecules
  correspond to sites with charge densities $0.5+\Delta n/2$ and
  $0.5-\Delta n/2$, respectively.}
\label{cofig}
\end{figure}
We first present a simple qualitative discussion of coupled
spin-singlet and CO formation at $\rho=\frac{1}{2}$. 
 The ideas are quite general and we argue
that the mechanism is {\it independent} of dimensionality.  The key
requirement is that the density must be exactly $\rho=\frac{1}{2}$, as
the effect requires commensurability.  Consider a single dimer of two
sites with one electron. The electron populations per site are 0.5
each, but the quantum mechanical wavefunction for the system is the
superposition $\frac{1}{\sqrt{2}}|10+01\rangle$, where $1$ and $0$ are
site charge densities. If one now brings two of these dimers together,
as in insert (a) of Fig.~\ref{cofig}, the composite wavefunction of
the two-dimer system can be written as
$\frac{1}{2}|1010+1001+0110+0101\rangle$. If the two electrons are in
a spin-singlet state then within the simple Hubbard Hamiltonian the
configuration $0110$, in which singlet stabilization can occur from a
single n.n. hop that creates a virtual double occupancy, must dominate
over the configurations $1010$ and $1001$,
in which singlet stabilization requires two and three hops,
respectively. Thus as the singlet bond between the dimers gets
stronger we expect a charge difference $\Delta n$ between sites
belonging to the same dimer (nominally between sites 1 and 2, or
between sites 3 and 4 in the linear chain of Fig.~\ref{cofig}).  While
some charge disproportionation must occur in finite linear chains from
end effects alone, we note that our proposed picture demands that
similar charge disproportionation occurs between members of the same
dimer even in the case of the periodic molecule shown in the insert
(b) of Fig.~\ref{cofig}.  In this case the charges on the sites
connected by the diagonal bond must be larger than 0.5, while the
charges on the two other sites must be smaller. Importantly, the
modulation of charge density, bond orders, as well as spin-singlet
pairing all occur cooperatively, and any of these observables may be
used as an order parameter in the case of a real transition.  In
Fig.~\ref{cofig} we have plotted $\Delta n$ versus the hopping
integral $t^\prime$ corresponding to the dotted bonds in the molecules
shown in the insert, for the ground spin-singlet state.  In both
cases, as $t^\prime$ increased from zero $\Delta n$ becomes nonzero
and increases with $t^\prime$.  Importantly, in the spin-triplet $S=1$
state the sign of $\Delta n$ is reversed in the cyclic molecule,
indicating repulsive interaction among the electrons.
\begin{figure}
\centerline{\resizebox{3.5in}{!}{\includegraphics{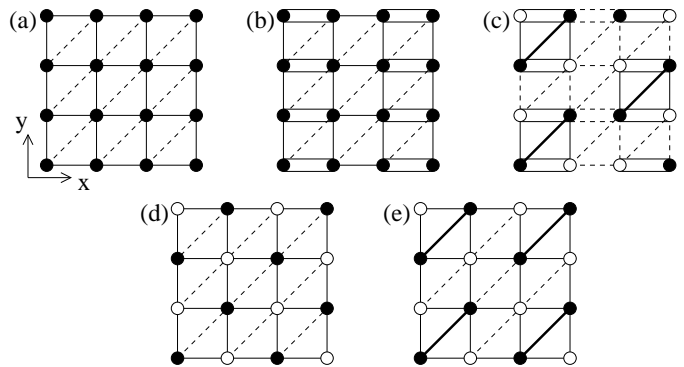}}}
\caption{(a) 2D lattice used for calculations in this paper, a square
  lattice with hopping $t_x=t_y\equiv t$ and frustrating bond
  $t_{x+y}\equiv t^\prime$ (dashed lines). (b) Dimerized lattice.
  Double (single) lines indicate stronger (weaker) bonds. Site charge
  densities are uniform in both (a) and (b).  (c) The PEC state as it
  occurs in this lattice.  Filled (open) circles correspond to sites
  with charge density $\rho=0.5+\delta$ ($\rho=0.5-\delta$)
  \cite{Li10a}.  Heavy line shows the location of singlet-paired
  sites.  (d) Wigner crystal charge ordering occurring for large $V_x$
  and $V_y$. (e) Wigner crystal-spin gap phase with bond alternation
  along diagonal directions. See section \ref{wcphase}.}
\label{lattices}
\end{figure}

Coupled CO and singlet formation occur also in the SP
state in 1D $\rho=\frac{1}{2}$ systems \cite{Ung94a,Clay03a,Clay07a}.
A key difference from the SP transition in $\rho=1$ is that for
$\rho=\frac{1}{2}$, a MI transition first occurs at
a intermediate temperature, followed by the SP transition at low
temperature.  For $\rho=\frac{1}{2}$, the MI leads to either a
bond-dimerized or to a charge-ordered WC state with equal bond lengths
\cite{Clay07a}.  Provided the n.n. Coulomb interaction is not too
strong, a SP transition occurs from either insulating state \cite{Clay07a},
 resulting
in a ground state with period-4 CO $\cdots$0110$\cdots$. 
 This state is the simplest realization of the PEC and
may visualized as a second dimerization of dimer units of molecules;
the singlet bond giving the spin gap (SG) forms {\it between} adjacent
dimer units.  The intermediate temperature bond-dimerized states, WC
state, and PEC state with SG are all found experimentally in quasi-1D
CTS \cite{Clay07a}.

Zigzag ladder systems, coupled two-stack systems in which each site on
one stack is coupled to two sites on the other stack, are a second
realization of the PEC state.  CTS zigzag ladder materials that are
$\rho=\frac{1}{2}$ have been found with spin-gap transition
temperatures much larger than in 1D $\rho=\frac{1}{2}$ SP materials
\cite{Rovira00a}.  The insulating ground state in this case may be
understood again as a PEC state occurring in a zigzag ladder lattice,
with singlet bonds oriented between the two chains \cite{Clay05a}.
Unlike the 1D case, bond orders are now modulated in several lattice
directions, leading to a larger SG than for the 1D PEC case
\cite{Clay05a}. Interestingly, in both the linear chain and the zigzag
ladder, the $\rho=\frac{1}{2}$ PEC is obtained by simply removing
alternate spin-singlet bonds from the corresponding $\rho=1$ VBS, and
replacing them with pairs of vacancies.

\subsection{Competing broken symmetries in 2D}

In this paper we will focus on ground state solutions of Eq.~\ref{ham}
in 2D in the presence of variable lattice frustration.  The lattice we
choose is a 2D square lattice with a single frustrating bond, as shown
in Fig.~\ref{lattices}(a).  Thus $\nu=\{x,y,x+y\}$ within
Eq.~\ref{ham} for this lattice.  In most of the results we will
present, $t_x=t_y\equiv t$, although we will also consider $t_x\neq
t_y$ in some cases.  Energies will be given in units of $t$.  We will
take the frustrating bond $t_{x+y}\equiv t^\prime$ in the range $0 \le
t^\prime < 1$, covering the wide region between the unfrustrated
square lattice ($t^\prime=0$) and the nearly isotropic triangular
lattice ($t^\prime=1$).  For the inter-site e-p coupling, unless
denoted otherwise we choose $\alpha_x=\alpha_y\equiv\alpha$ and
$\alpha^\prime=0$, with similarly identical spring constants
$K^x_\alpha=K^y_\alpha\equiv K_\alpha$. For all calculations we assume
periodic boundary conditions.

Our calculations are largely for $V_x=V_y$, but variable
$V_{x+y}\equiv V^\prime$.  In Reference \onlinecite{Li10a} we
presented limited numerical results for a select set of Coulomb
interaction parameters ($U=4$, $V_x=V_y=1$, $V^\prime=0$)
demonstrating transition from N\'eel antiferromagnetism to the PEC
state in this lattice when $t^\prime$ exceeds a critical value
$t^\prime_c$.  For completeness and for giving an introduction to the
various competing states in 2D we briefly review these results
here. For small $t^\prime$, the self-consistent solution of
Eq.~\ref{ham} gives spontaneous dimerization along the $x$ axis as
shown in Fig.~\ref{lattices}(b). The dimerized lattice state is
effectively $\frac{1}{2}$-filled with one carrier per dimer and has
N\'eel AFM order between dimers for finite $U$.
The N\'eel order is very clearly observable from the spin-spin
correlations calculated for $4\times 4$ clusters \cite{Li10a}.
Importantly, the charge density $\langle n_i \rangle$ of all sites is
exactly 0.5.

As $t^\prime$ increases, frustration reduces the strength of the
antiferromagnetic correlations. Provided the n.n. Coulomb interaction
$V_{ij}$ is not too strong (see below), at a critical
$t^\prime=t^\prime_c$ antiferromagnetic correlations disappear and
charge disproportionation develops, with the charge densities within
each dimer becoming inequivalent. This is shown in
Fig.~\ref{lattices}(c).  In this state, the charge densities follow
the pattern $\cdots$1100$\cdots$ along the $x$ and $x+y$ directions,
and the pattern $\cdots$1010$\cdots$ along the $y$ direction.  The
bond distortion is period-4 along $x$ and period-2 along $y$.  The
strongest bond orders $\langle B_{ij} \rangle$ occur between adjacent
charge-rich `1--1' sites in the $x+y$ direction.  Importantly, the
spin-spin correlations change dramatically for
$t^\prime>t^\prime_c$. They are strongly negative between the bonded
`1--1' sites along the $x+y$ direction (see Fig.~\ref{lattices}(c)),
and are nearly zero between either member of the pair and all other
sites, indicating the formation of spin-singlet bonds \cite{Li10a}.
Any of the observables $\Delta n$, bond order between the charge-rich
sites, or z-z spin correlations, $\langle S^z_iS^z_j\rangle$, may be
used as order parameters for the PEC state \cite{Li10a}.

In section \ref{results} we present further details of the PEC phase
and the full parameter dependence of Eq.~\ref{ham}, with the goal of
demonstrating that {\it (i) the transition to the PEC that we are
  interested in is driven by quantum effects due to frustration only;
  and (ii) the PEC occurs over a broad region of parameter space.}
Given the number of parameters in Eq.~\ref{ham}, it should be
relatively easy to generate CO driven by specific (presumably
artificial) choices of $V_{ij}$. Such classical results would be
uninteresting.  We therefore consider several distinct choices of
Coulomb interactions: (i) $U>0$ and all $V_{ij}=0$, (ii) $U>0$,
$V_x=V_y=V$, and $V^\prime=0$, (iii) $U>0$ and $V_x=V_y=V^\prime=V$.
PEC formation occurs in all of these parameter regions. We also show that
for sufficiently strong n.n. Coulomb interactions corresponding to
parameter region (ii), the WC phase with checkerboard CO
(Fig.~\ref{lattices}(d)-(e)) is the ground state of Eq.~\ref{ham}.  We
also will consider other modifications of the basic lattice, viz.,
sign of $t^\prime$ opposite site to $t$, and
$t_x\neq t_y$. Finally, we will argue that our
results are not consequences of finite size effects and are to be
expected in the thermodynamic limit.

\section{Results}
\label{results}

\subsection{ $U>0$, $V_x=V_y=V^\prime=0$}
\label{results-v0}

\begin{figure*}
\centerline{\resizebox{7.0in}{!}{\includegraphics{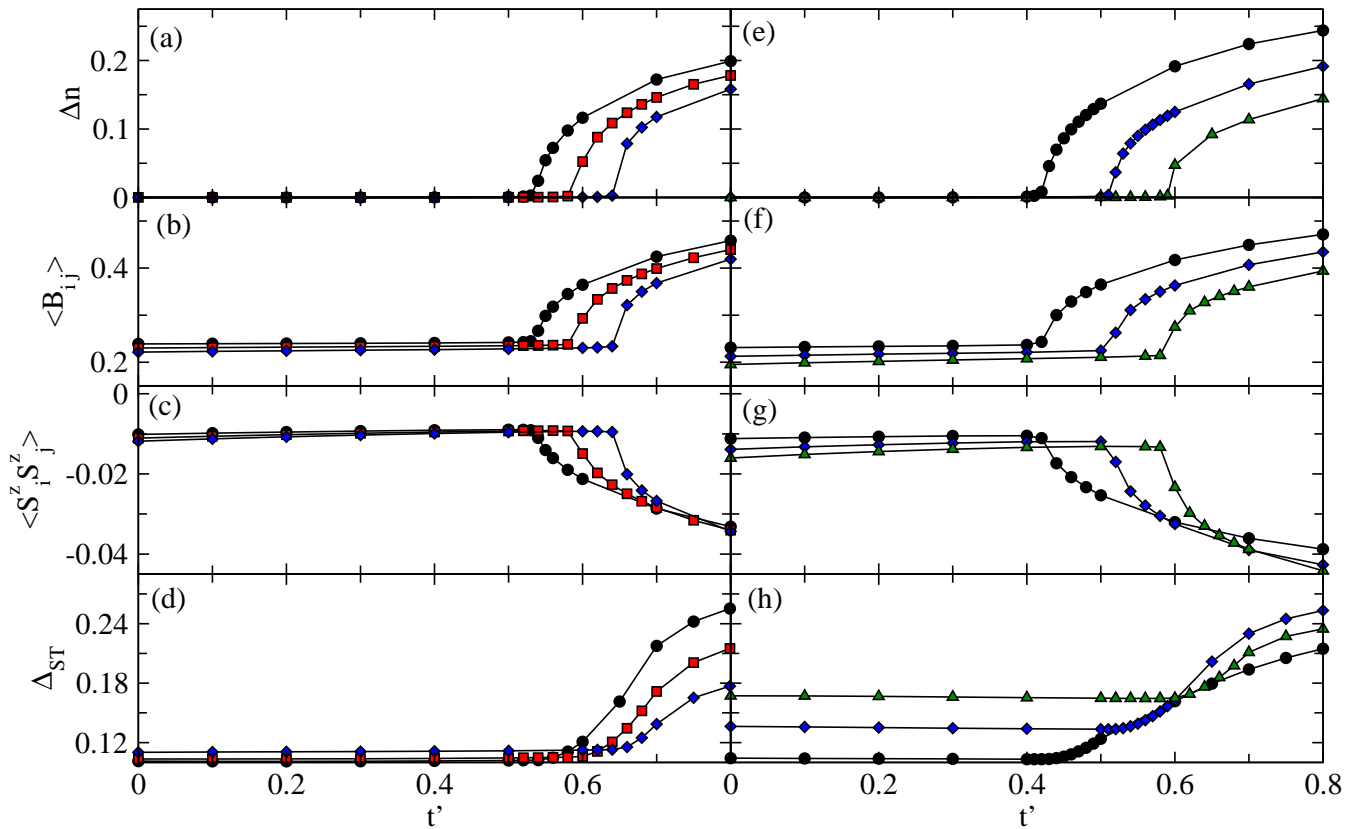}}}
\caption{(Color online) Order parameters for the 4$\times$4 lattice
  versus $t^\prime$. Parameters are $\alpha=1.1$, $\beta=0.1$, and
  $K_\alpha$=$K_\beta$=2.  
 Circles, squares, diamonds,
  and triangles are for $U=$ 2, 3, 4, and 6, respectively.
$V_x=V_y=V^\prime=0$ for (a)--(d), and
  $V_x=V_y=1$ and $V^\prime=0$ for (e)--(h). (a) and (e) show the
  charge disproportionation $\Delta n$, (b) and (f) the bond order
  $B_{i,j}$ between charge-rich sites $i$ and $j$ connected by
  $t^\prime$ (see Fig.~\ref{lattices}(c)), (c) and (g) spin-spin
  correlations between these sites, and (d) and (h) the
  singlet-triplet gap $\Delta_{\rm{ST}}$.  
In all cases the CO pattern is as shown in Fig.~\ref{lattices}(c).
For all plots lines   are guides to the eye.}
\label{bigplot}
\end{figure*}

Figs.~\ref{bigplot}(a)-(d) show the charge disproportionation $\Delta
n$, bond order $\langle B_{ij}\rangle$,  n.n.  z-z spin-spin
correlation $\langle S^z_iS^z_j\rangle$,
and spin gap $\Delta_{ST}$ 
 as a function of $t^\prime$. $\Delta_{ST}$ is defined as the excitation energy from the
ground $S=0$ state to the lowest $S=1$ state.
$U$ here is finite but all $V$ terms are zero.  Sites $i$ and $j$
in Figs.~\ref{bigplot}(b)-(c)
correspond to two `1' sites in the PEC state connected by a $t^\prime$
bond (filled circles connected by heavy lines in
Fig.~\ref{lattices}(c)).  As with the results in Reference
\onlinecite{Li10a} which included the $V_x$ and $V_y$ interactions,
for small $t^\prime$ the charge density is 0.5 on all sites, and
antiferromagnetic order can be seen in the spin-spin correlations (not
shown here, see Fig.~3(a) in Reference \onlinecite{Li10a}).
Nonzero $t^\prime_c$ in Fig.~\ref{bigplot} is  a consequence of the
nature of the diagonal bonds in Fig.~\ref{lattices}(b); 
the diagonal bonds inside each plaquette  with two strong dimer
bonds actually strengthen the AFM, and only the inter-plaquette
diagonal bonds have a frustrating effect. At small $t^\prime$ these two
effects appear to cancel, and there is only a weak effect on the AFM.

For $t^\prime>t^\prime_c$, $\Delta n$ becomes nonzero, with the charge
pattern as shown in Fig.~\ref{lattices}(c). Similarly, the bond order
between paired `1--1' sites increases abruptly, and the z-z spin-spin
correlation between these sites becomes strongly negative, and nearly
zero with all other lattice sites (see Reference \onlinecite{Li10a}),
indicating formation of a singlet bond.  Although $\Delta_{ST}$ is
nonzero in all cases in a finite cluster, we nevertheless see a large
jump in $\Delta_{ST}$ at $t^\prime_c$, also indicating spin-singlet
formation.  The increase in bond order strength and strength of
spin-spin correlation clearly follow the same pattern as  $\Delta
n$.

 These results show that the n.n. Coulomb interaction is not essential
 for formation of the PEC state.  Unlike the WC phase where CO is
 driven by the n.n. Coulomb interaction, the PEC state is a
 consequence of geometric lattice frustration.  Increasing $U$ moves
 $t^\prime_c$ to larger $t^\prime$: $U$ tends to strengthen the AFM
 phase and therefore increasing $U$ is expected to make the AFM order
 persist for stronger lattice frustration.  The transition also
 becomes more discontinuous to changes in $t^\prime$ as $U$ increases,
 suggesting it may be continuous for small $U$ and first order for
 large $U$.

In both the 1D and in the zigzag ladder lattice, the PEC state occurs
unconditionally even in the noninteracting limit ($U=V_{ij}=0$) for
any finite e-p coupling.  In these two cases, the unconditional occurrence
of PECs is a consequence of simple nesting.  In contrast, in the
isotropic $\rho=\frac{1}{2}$ 2D band considered here, the lack of
nesting forbids an unconditional Peierls transition.
In agreement with this, we found that for $U\alt1$, the PEC phase did
not occur. Instead, the self-consistent calculations converged to
disordered states with no clear charge pattern. This is an indication
that in the thermodynamic limit, the preferred ground state is one of
uniform charge.  This result is reminiscent of that in quantum spin
systems:  the simplest VBS transition, the SP transition, can be
predicted from nesting behavior in the 1D XY model following
Jordan-Wigner transformation \cite{Beni72a}. This is, however, not
true in 2D frustrated spin systems.

\subsection{$U>0$, $V_x=V_y>0$, $V^\prime=0$}
\label{wcphase}

\begin{figure}
\centerline{\resizebox{3.3in}{!}{\includegraphics{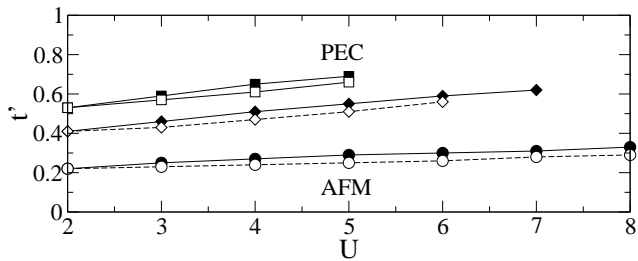}}}
\caption{Phase diagram as a function of $t^\prime$ and $U$, for
  $\beta=0.1$, and $K_\alpha=K_\beta=2$. Squares are the AFM--PEC
  phase boundary for $\alpha=1.1$ and $V_x=V_y=V^\prime=0$; Diamonds
  are for $\alpha=1.1$, $V_x=V_y=1$, and $V^\prime=0$; Circles are for
  $\alpha=1.2$, $V_x=V_y=1$, and $V^\prime=0$.  For each case, filled
  (open) points correspond to positive (negative) $t^\prime$. Lines
  are guides to the eye.}
\label{uv-phase}
\end{figure}

We next consider the effect of n.n. Coulomb interactions
$V_x=V_y=V$, but $V^\prime=0$.  Figs.~\ref{bigplot}(e)-(h) show
the same order parameters as in Figs.~\ref{bigplot}(a)-(d) for
$V=1$. Comparing the data with and without $V$,
 the effect of moderate $V$ is to {\it strengthen}
the PEC state---the magnitude of all order parameters increase when
$V>0$ for a fixed value of $U$.  For fixed $U$, the AFM--PEC boundary
$t^\prime_c$ also moves to smaller $t^\prime$ with increasing $V$.
Fig.~\ref{uv-phase} shows the phase diagram in the $t^\prime$-$U$
plane for both the $V=0$ and $V>0$ cases.  Here, for each value of
$U$, $t^\prime$ was varied until the AFM--PEC transition occurred, and
the first $t^\prime$ where $\Delta n$ became nonzero was taken as the
AFM--PEC boundary.

One expects that when $V$ is above a critical value $V_c$, the
$\cdots$1100$\cdots$ PEC CO will give way to the checkerboard WC
state. In 1D, this transition occurs exactly at $V_c=2$ in the limit
$U\rightarrow\infty$, and at a larger $V_c$ for finite $U$
\cite{Penc94a,Clay03a}.  Previous exact diagonalization for a 2D
cluster\cite{Merino05a} (without however e-p interactions or
$t^\prime$ as considered here) found $V_c\approx 2.1$ for $U=10$, and
showed that $V_c$ increases when $V^\prime>0$.

\begin{figure}
\centerline{\resizebox{3.3in}{!}{\includegraphics{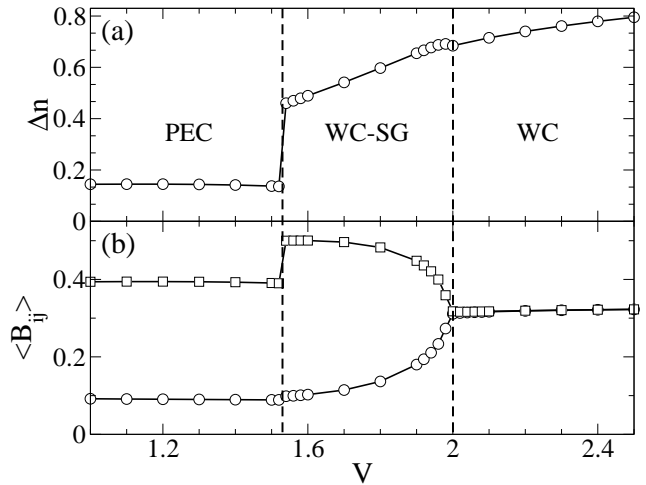}}}
\caption{Variation of order parameters with $V=V_x=V_y$ ($V^\prime=0$)
  for $U=6$, $t^\prime=0.8$, $\alpha=1.1$, $K_\alpha=K_\beta=2$,
  $\beta=0.1$.  (a) charge disproportionation $\Delta n$ (b) bond
  orders along two successive $t^\prime$ bonds. In PEC region, these
  correspond to bonds between sites with `1--1' (squares) and `1--0'
  (circles) occupancy.  In the WC-SG and WC regions, both bonds are
  between `1--1' sites.  Lines are guides to the eye.}
\label{wcplot}
\end{figure}

Fig.~\ref{wcplot} shows the evolution of $\Delta$n and diagonal bond
orders with $V$ for $U=6$.  While difficult to see in
Fig.~\ref{wcplot} due to the choice of axis scales and parameters, in
the PEC region $\Delta n$  increases with increasing $V$.  For the
parameters of Fig.~\ref{wcplot} ($U$=6, $t^\prime=0.8$, $\alpha=1.1$,
$\beta=0.1$, $K_\alpha=K_\beta=2$) the charge order pattern changes
from the PEC (Fig.~\ref{lattices}(c)) to the checkerboard WC
(Fig.~\ref{lattices}(d)-(e)) at $V\approx1.52$. In addition, $\Delta
n$ increases sharply when entering the WC phase. At larger
$V\approx2.0$ a slight cusp occurs in the $\Delta n$ versus $V$ plot.
At the same time, the pattern of bond orders changes: for $V<2$ in the
WC phase the bond orders alternate strong-weak along the $x+y$
direction (as shown in Fig.~\ref{lattices}(e)), while for $V>2$ the
bond orders along $x+y$ are uniform (Fig.~\ref{lattices}(d)).  Within
the WC phase region there are therefore two sub-phases: a phase with equal
length bonds in the diagonal $x+y$ ($t^\prime$) directions, and a
phase in which these bonds become dimerized. The added bond
dimerization in the diagonal direction will result in a spin gap, and
we denote this phase as the Wigner Crystal--Spin gap (WC-SG) phase.
A similar spin-gapped WC phase can be found in a small
region of parameter space in the 1D model \cite{Kuwabara03a,Clay03a}.
\begin{figure}
\centerline{\resizebox{3.3in}{!}{\includegraphics{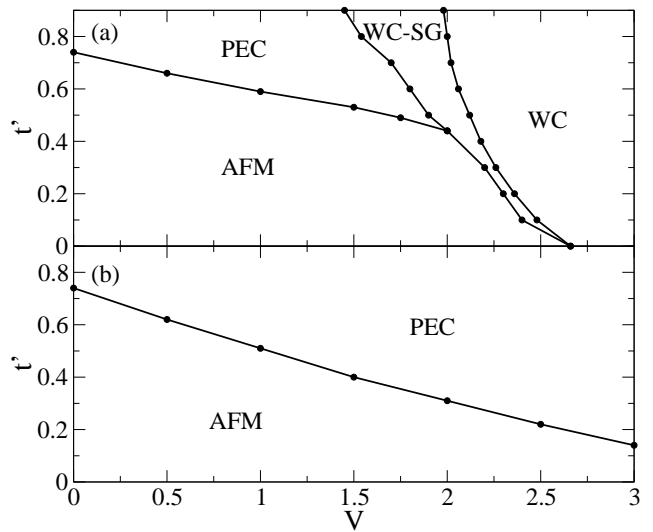}}}
\caption{(a) Phase diagram for the 4$\times$4 lattice as a function of
  $t^\prime$ and $V=V_x=V_y$, $V^\prime=0$, for $U=6$, $\alpha=1.1$,
  $\beta=0.1$, and $K_\alpha=K_\beta=2.0$.  (b) Same as (a), but with
  $V=V_x=V_y=V^\prime$. For both (a) and (b), points between
  antiferromagnetic and PEC phases are determined as discussed in
  section \ref{afm-pec}; The boundary between PEC, WC-SG, and WC
  phases is discussed in the text. Lines are guides to the eye.}
\label{vt-phase}
\end{figure}

Fig.~\ref{vt-phase}(a) shows the resulting phase diagram in the
$t^\prime$-$V$ plane for $U$=6. The $V_c$ we find in the $t^\prime=0$
limit is slightly larger ($U=6$, $V_c\approx2.6$) than the results of
Reference \onlinecite{Merino05a} ($U=10$, $V_c\approx2.1$); however, both the
smaller $U$ here as well as the e-p coupling in Eq.~\ref{ham} would be
expected to increase $V_c$.  Unlike the AFM, PEC, and WC phases, the
WC-SG phase is limited to a relatively narrow range of
parameters. Larger lattice calculations are needed to confirm whether
the WC-SG phases persists in the thermodynamic limit.

\subsection{$U>0$, $V_x=V_y=V^\prime>0$ }

The $V^\prime$ interaction destabilizes the checkerboard-pattern WC,
leading to a metallic phase in the absence of e-p interactions
\cite{Merino05a}.  Here we will consider parameters
$V_x=V_y=V^\prime=V$.  For Eq.\ref{ham} without e-p interactions,
$U=10$, and $t^\prime\leq 0.1$, exact diagonalization
found\cite{Merino05a} that in this case a metallic phase exists for
$V$ up to at least $V=$5. Charge fluctuations within the metallic
phase adjacent to the WC were speculated to cause a CO-to-SC
transition \cite{Merino01a}. In our calculations, summarized in the
phase diagram in Fig.~\ref{vt-phase}(b) for $U=6$, we also found that
the WC phase does not occur, but rather than being metallic the system
is insulating---either AFM at small frustration or PEC at large
frustration.  In this case the AFM--PEC transition can occur over a
wide range of lattice frustration, $0.2\alt t^\prime_c\alt 0.7$.
Within the PEC phase the CO pattern remains the same for all $V$,
although $\Delta n$ increases with $V$ as in the $V^\prime=0$ case
considered in the previous section.  Although our calculations are for
one value of $V^\prime$ only, Fig.~\ref{vt-phase}(b) suggests that the
PEC region is broadened relative to that in Fig.~\ref{vt-phase}(a) for
any $V^\prime\neq 0$.

\subsection{Bandstructure and electron-phonon coupling}

As shown in the previous section, variation of Coulomb interactions
can cause a substantial variation in the extent of frustration needed
to form the PEC state. Next we show the effect of varying the
one-electron parameters in Eq.~\ref{ham}, $t_\nu$ and $\alpha_\nu$.

Due to the lack of particle-hole symmetry in the anisotropic
triangular lattice, differences might be expected when $t^\prime$ is
taken as negative.  However, as Fig.~\ref{uv-phase} shows, we found
only a small variation in the AFM--PEC phase diagram when the sign of
$t^\prime$ is changed.  This is consistent with the expected mechanism
for spin frustration in an effectively $\frac{1}{2}$-filled band: the
frustrating exchange interaction is proportional to $(t^\prime)^2$, so
reversing the sign of $t^\prime$ should only change the effective
frustration at higher order.  
\begin{figure}
\centerline{\resizebox{3.3in}{!}{\includegraphics{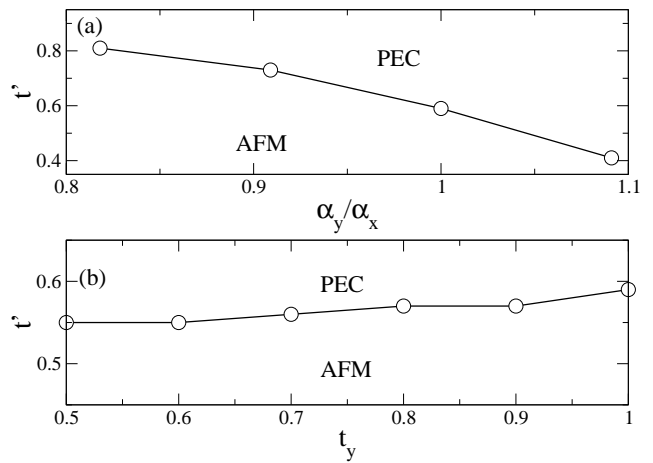}}}
\caption{Phase diagram variation on hopping and electron-phonon
interaction anisotropy.
In both (a) and (b),  $U=6$, $V_x=V_y=1$, $V^\prime=0$, 
 $\beta=0.1$, and $K_\alpha=K_\beta=2$. 
In (a), $t_x=t_y=1$ and $\alpha_x=1.1$, but $\alpha_y$ is varied.
In (b), $t_x=1$ and $t_y$ is varied, with identical $\alpha_x=\alpha_y=1.1$.
Lines are guides to the eye.}
\label{txty-alpha-phase}
\end{figure}

Fig.~\ref{uv-phase} also shows the effect of changing $\alpha$.  
As expected for a cooperative transition,
stronger
e-p coupling increases the size of the PEC region.  The effect of
anisotropy of the inter-site e-p interaction is shown in
Fig.~\ref{txty-alpha-phase}(a), where $\alpha_x$ is fixed at 1.1 and
$\alpha_y$ is varied.  Increasing either $\alpha_x$ or $\alpha_y$
separately strengthens the PEC, as shown in
Fig.~\ref{txty-alpha-phase}(a), where the $\alpha_y/\alpha_x$ ratio is
varied.  The AFM--PEC phase diagram is relatively insensitive to the
ratio of $t_y/t_x$. In Fig.~\ref{txty-alpha-phase}(b), $t_x$ is fixed
at 1 and the value of $t_y$ is varied---the resulting $t^\prime_c$
does not depend sensitively on the choice of $t_y/t_x$.

Summarizing sections \ref{results}(a)-(d), we see the PEC state in our
calculations for a wide range of parameters. In every case, we first
arrived at parameters that placed the system in the AFM or WC phases
in the $t^\prime=0$ limit, and then varied only $t^\prime$.  Thus the
transition to PEC is a consequence of frustration alone.  With
variation of e-p coupling and Coulomb interactions, the amount of
frustration needed to drive the AFM--PEC transition can vary over a
sizable range of frustration.
We will discuss the implications of this further in relationship to the
CTS materials further in section \ref{discussion}.

\subsection{Finite-size issues}

We have performed several checks on our calculations that indicate that the
PEC state found in our numerical calculations is an intrinsic property
of Eq.~\ref{ham} and {\it not} induced by finite size effects.

{\it (i) Noninteracting bandstructure:} One common finite-size effect
in numerical calculations are changes in the Fermi-level degeneracy or
level crossings of the noninteracting system.  In the lattice of
Fig.~\ref{lattices}(a) the Fermi level degeneracy does not change
throughout the range $0<t^\prime<1$, remaining 2-fold degenerate
throughout this range. This degeneracy is broken by $x$-axis
dimerization, shown in Fig.~\ref{lattices}(b), giving a nondegenerate
Fermi level for $0<t^\prime<1$.  At $t^\prime=1$ the degeneracy at the
Fermi level increases to 6-fold; in the presence of interactions the
ground state can become triplet ($S$=1) for $t^\prime\agt0.8$.  Hence
we stop at $t^\prime\alt0.8$ where the ground state is $S=0$.

{\it (ii) Interactions:} As mentioned above, no transition occurs for
$U=V_{ij}=0$. This further indicates that the transition is not
a feature of the single-particle bandstructure.

{\it (iii) Commensurability:} We have verified that the PEC state does
not occur for electron densities different from $\rho=\frac{1}{2}$;
for example, no transition to PEC or any ``nonmetallic'' state occurs
for 6 or 10 electrons on the $4\times4$ lattice.

\section{Application to real materials}
\label{discussion}

In the following we discuss how our theory applies to real materials,
and may even give insight to the mechanism of correlated-electron
superconductivity in $\rho=\frac{1}{2}$ materials.

\subsection{Application to organic CTS}

The superconducting organic CTS share many characteristics of other
strongly-correlated superconductors, in particular the high-T$_c$
cuprates, including reduced dimensionality and the presence of AFM
near SC. At the same time, SC in the CTS occurs under pressure at a
constant carrier density of $\rho=\frac{1}{2}$ rather than under the
influence of doping. A variety of exotic insulating states {\it in
  addition to AFM}, including CO \cite{Takahashi06a} and spin-gapped
states \cite{Mori06a,Tamura09a}, as well as possible QSL states \cite{Kanoda11a}, are
proximate to the superconducting state in the CTS. 
Our work shows that only the AFM phase is described by the
dimer Mott-Hubbard model.
Since in all cases
the experimental systems are structurally related, with identical or
near-identical molecular components, we believe that the same
mechanism of SC should apply to them.  In Reference \onlinecite{Li10a} we
had pointed out how the PEC concept can perhaps lead to such a
unifying theory. Here we expand on this theme.

\subsubsection{CTS with PEC insulating states}

We briefly review here experimental
evidence for PEC formation in several 2D CTS families.

{\it (i) $\theta$-(ET)$_2$X:} CO corresponding to the PEC and spin gap
are found in the $\theta$-(ET)$_2$MM$^\prime$(SCN)$_4$ family
\cite{Mori06a}.  In $MM^\prime$=RbZn, the CO occurs below the MI
transition at $T\sim$190K, while the SG appears below 20K
\cite{Mori06a}.  The charge order pattern in the CO phase below the MI
transition has been experimentally determined for $MM^\prime$=RbZn and
follows a horizontal stripe pattern (see Fig.~6 in Reference
\onlinecite{Watanabe04a}). The horizontal CO is definitely not the
WC. Rather, the CO pattern is precisely as expected in the PEC, with
$\cdots$1100$\cdots$ CO along the two directions of largest hopping
(the $p$-directions in the $\theta$-(ET)$_2$X lattice), and
$\cdots$1010$\cdots$ order along the direction of weakest hopping
($c$-direction in $\theta$-(ET)$_2$X). Experiments have revealed that
with decreasing temperature, the $c$-axis lattice parameter decreases
\cite{Watanabe07a}. The decrease in the lattice parameter implies
increased carrier hopping in this direction and therefore increased
frustration within our theory, giving the transition to the singlet
PEC and observed spin gap.
\smallskip

{\it (ii) $\alpha$-(ET)$_2$X:} The crystal structure of
$\alpha$-(ET)$_2$X is quite similar to that of $\theta$-(ET)$_2$X. The
existence of a SG opening below 136K has been known for some time in
$\alpha$-(ET)$_2$I$_3$ \cite{Rothaemel86a}. Below the 136K transition
CO is found which has been confirmed to be of the same pattern as in
$\theta$-(ET)$_2$X; see for example Fig.~2 in Reference
\onlinecite{Tajima09a}.
\smallskip

{\it (iii) $\beta$-(meso-DMET)$_2$PF$_6$:} This 2D CTS exhibits a
pressure-induced transition from CO to SC \cite{Kimura04a,Kimura06a}.
While the charge order pattern in this CTS is referred to as
``checkerboard'' by the authors, the checkerboard pattern refers to
meso-DMET {\it dimers} as units. In terms of meso-DMET {\it monomer}
units, the CO pattern is the same as the PEC in
Fig.~\ref{lattices}(c), with $\cdots$1100$\cdots$ in two directions
and $\cdots$1010$\cdots$ along the third direction (see Fig.~2 in
Reference \onlinecite{Kimura06a}.)
\smallskip

{\it (iv) $\beta^\prime$-X[Pd(dmit)$_2$]$_2$:} In this family the
materials dmit molecules are arranged in dimers.  The frustration
varies with the cation $X$, with the least frustrated in the series
showing AFM order \cite{Kato04a}. Among the materials with larger
frustration, $X=EtMe_3P$ has a SG transition at 25K to what has been
described as a VBS state \cite{Tamura06a}. The experimentally
determined bond and charge distortion patterns (see Fig.~3(b) in
Reference \onlinecite{Tamura06a}) are exactly as expected for the PEC,
with period 4 charge and bond distortions along what is the $x$ axis
in Fig.~\ref{lattices}. The intradimer charge disproportionation in
particular argues against the SG state from being a simple VBS, which
would require equal charge densities on all the molecules.

\subsubsection{$\kappa$-(ET)$_2$Cu$_2$(CN)$_3$ and EtMe$_3$Sb[Pd(dmit)$_2$]$_2$:
QSL or charge-disproportionated states?}

There has been much recent interest in these CTS \cite{Kanoda11a}
specifically because they present possible realizations of the long
awaited QSL
\cite{Lee05a,Motrunich05a,Galitski07a,Lee07a,Lee08a,Qi08a,Qi09a,Xu09a,Grover10a}.
In both cases the materials have nearly isotropic triangular lattices
of dimer unit cells (corresponding to Fig.~\ref{lattices}(b) with
$t^\prime=1$) within an effective $\rho=1$ model.
 In $\kappa$-(ET)$_2$Cu$_2$(CN)$_3$ (hereafter $\kappa$-CN) the estimate
for the Heisenberg exchange integral between n.n. dimers 
is $J\sim$220--250 K \cite{Shimizu03a,Zheng05a}.  $^1$H
NMR experiments find absence of long range magnetic order down to 32
mK \cite{Shimizu03a}.  Very similar behavior is also seen in
EtMe$_3$Sb[Pd(dmit)$_2$]$_2$ (hereafter dmit-Sb) \cite{Itou10a}. The
insulating ground state in $\kappa$-CN is close to SC, transition to
superconductivity occurring under moderate pressure
\cite{Komatsu96a}. SC is found in the dmit family as well
\cite{Shimizu07a}.

Recent experiments in both $\kappa$-CN and dmit-Sb have found
peculiarities that appear to be unexpected within QSL theories.  Below
we list experiments that seem to indicate that apparent QSL behavior
at low temperatures is giving way to a ``hidden order'', and perhaps
even charge disproportionation, which by itself would be against
spin-only models.
\smallskip

(i) A second order phase transition is seen at 6 K in $\kappa$-CN in
measurements of heat-capacity C$_p$ \cite{Yamashita08a}, $^{13}$C NMR
relaxation rate 1/T$_1$ \cite{Shimizu06a}, and lattice expansion
coefficients \cite{Manna10a}. The last experiment finds strong lattice
effects at the transition, indicating possible role of charge degrees
of freedom \cite{Manna10a}. A symmetry-breaking and/or topological
ordering transition at T $<$ 1 K has also been observed in dmit-Sb
\cite{Itou10a}.  
\smallskip

(ii) The specific heat C$_p$ in $\kappa$-CN is linear in T for T
between 0.75 and 2.5 K, indicating a gapless energy spectrum
\cite{Yamashita08a}.  The Sommerfeld coefficient $\gamma$ is nonzero
and large even at $T=$75 mK \cite{Yamashita08a}.  Equally
perplexingly, C$_p$ is independent of magnetic field up to 8 T,
indicating absence of Zeeman coupling of spins to the field
\cite{Yamashita08a}.  In contrast, thermal conductivity measurements
down to 80 mK indicate a spin gap \cite{Yamashita09a}.
\smallskip

(iii) The temperature dependence of the thermal
conductivity of dmit-Sb suggests gapless mobile excitations, but
magnetic field dependence of the thermal conductivity again indicates
a gap \cite{Yamashita10a}.  
Taken together, (ii) and (iii) suggest
gapless spin-singlet excitations but gapped spin-triplet excitations
in $\kappa$-CN and perhaps also dmit-Sb.
It is conceivable that the gap in dmit-Sb is nodal \cite{Yamashita10a}.
\smallskip

(iv) Measurements of dielectric response in $\kappa$-CN have shown
increasing and frequency-dependent dielectric constant below 60 K, and
possible antiferroelectric ordering of dipoles at T$_c\sim$6 K
\cite{AbdelJawad10a}. The latter requires unequal site charges
on the molecules within the dimer unit cells \cite{AbdelJawad10a,Seo10a}. It
may be relevant in this context that $^{13}$C-NMR experiments on both
$\kappa$-CN \cite{Shimizu06a} and dmit-Sb \cite{Itou08a} find unusual
line broadenings at low T that cannot be ascribed to disorder
\cite{Gregor08a}, which might also indicate charge
disproportionation. Similar line broadening at low T in
EtMe$_3$P[(dmit)$_2$]$_2$ occurs at the transition to the PEC
\cite{Tamura06a}, which in turn gives way to superconductivity under
pressure \cite{Shimizu07a}.
\smallskip

Although it is as yet not entirely clear whether or not existing 
spin-liquid models \cite{Lee05a,Motrunich05a,Galitski07a,Lee07a,Lee08a,Qi08a,Qi09a,Xu09a,Grover10a} with appropriate modifications can explain the 
above anomalies, it appears that for both $\kappa$-CN and dmit-Sb, spin 
degrees of freedom alone cannot describe the low temperature (below 6K in 
$\kappa$-CN and below 1K in dmit-Sb) properties. Rather, any description 
of the ground state must involve charge as well as spin degrees of 
freedom. Assuming that the above experiments and their interpretations are 
correct, the proper theoretical model should have built-in significant 
electron-lattice coupling and should lead to charge disproportionation and 
excitation energy spectrum with gapless singlet excitations and gapped 
spin excitations. We believe that the highly frustrated $\rho=\frac{1}{2}$ 
model satisfies all of the above criteria, in addition to providing the 
starting point for a theory of superconductivity in the CTS (see below). 
T-linear specific heat but gapped magnetic susceptibility were noted in 
nonmetallic vanadium bronzes\cite{Chakraverty78a} as far back as 1978, 
and at the time was considered to be a distinctive proof for bipolarons. 
The nonzero spin-singlet degeneracy within this model comes from tunneling 
motion of the bipolarons, causing them to ``flip flop'' between 
equivalent configurations \cite{Chakraverty78a}. We propose that a 
similar mechanism is at play in the present case at large $t^\prime$. 
Particularly in $\kappa$-CN the lattice structure and the orientations of 
molecules are such that the spin-singlet bonds, which occur between 
monomers belonging to two neighboring dimers can flip flop between the 
monomers (see Fig.~4 in Reference \onlinecite{Li10a}.) Within our proposed picture for 
$\kappa$-CN, the transition at 6K is to the spin-singlet 
charge-disproportionated state, with short range fluctuating order. 
Magnetic excitations even in this state, however, require breaking the 
spin-singlet bonds. Note that the observed difference between $\kappa$-CN 
and dmit-Sb is expected within the $\rho=\frac{1}{2}$ model, since the 
simple description as a triangular lattice of dimers is no longer enough 
and the detailed couplings between the monomers in the materials are 
indeed different because of their different crystal structures.

\subsubsection{Consequence of stronger frustration---paired electron
liquid and superconductivity}

As mentioned above, numerical results for $t^\prime\agt 0.8$ are not
useful due to the highly degenerate ground state becoming spin-triplet
in our finite clusters.  We have suggested elsewhere that the occupied
spin-singlet bonded `1--1' sites of the PEC can be thought of as {\it
  effective} single sites doubly occupied by charge carriers, and
similarly the pairs of `0--0' vacancies can be thought of as single
vacant sites \cite{Mazumdar08a}. Such a mapping would transform the
PEC to an effective checkerboard CO with alternate sites (in the
square lattice representation) occupied by double occupancies and
vacancies. The effective Hamiltonian that describes the checkerboard
CO in this case is a $\rho=1$ extended Hubbard model with weak {\it
  attractive} $U$ whose origin is the exchange interaction that
stabilizes the singlet bond within the original $\rho=\frac{1}{2}$
Hamiltonian.  The n.n. interaction within the effective Hamiltonian
remains repulsive to simulate the checkerboard ordering of the
effective doubly occupied sites.

While such a mapping is not rigorous, similar mapping of
n.n. spin-bonded sites to double occupancies has been routinely used
in the literature on bipolaron models
\cite{Alexandrov81a,Micnas90a}. The difference between our work and
traditional bipolaron models is that the spin-singlet bonding within
our work is driven primarily by AFM correlations, while within the
bipolaron models it is a consequence of effective attraction due to
the overscreening of the e-e repulsion by e-p interactions
\cite{Micnas90a,Alexandrov94a}.  We have investigated the consequences
of stronger frustration\cite{frustration-note} in the
$\rho=\frac{1}{2}$ anisotropic triangular lattice within the effective
$\rho=1$ extended Hubbard model with attractive $U$
\cite{Mazumdar08a}.  We found a CO-to-SC transition within the
effective model \cite{Mazumdar08a}, suggesting a PEC-to-SC transition
within the $\rho=\frac{1}{2}$ model with repulsive
interactions. Although this result is not a proof of transition to SC
within the repulsive $\rho=\frac{1}{2}$ model, it nevertheless is
instructive and provides the direction for future research.

Demonstration of SC within the actual model will require further
work. What is interesting though is that the proposed scenario can
give a unified approach to SC in all organic CTS, irrespective of
whether the insulating state proximate to SC is AFM \cite{Kanoda06a},
CO \cite{Mori06a,Tajima06a} or VBS \cite{Shimizu07a}. Recall that
within existing mean field theories the AFM-to-SC transition is driven
by spin fluctuations
\cite{Vojta99a,Schmalian98a,Kino98a,Kondo98a,Powell0507,Baskaran03a,Gan05a,Gan06a,Sahebsara06a,Watanabe06a,Kyung06a},
while the CO-to-SC transition is driven by charge fluctuations
\cite{Merino01a}. Even if we ignore that recent precise numerical
calculations \cite{Mizusaki06a,Clay08a,Tocchio09a} have demonstrated the
absence of SC within the proposed spin-fluctuation models in this
context (thereby raising doubts also about mean-field theory of
charge-fluctuation mediated superconductivity), different mechanisms
of SC for structurally related materials with identical or
near-identical molecular components appear to be unrealistic. Within
our proposed scenario, the charge-ordered or VBS systems that exhibit
SC are PECs in the semiconducting state, while the AFM systems under
pressure transform to PECs first and then to SC (although due to
actual crystal structures the ``width'' of the PEC region could be
narrow to vanishing within the latter). The SC phase within this scenario is a
paired-electron liquid. There is considerable overlap
between these ideas and the one proposed by Moulopoulos and Ashcroft
for the continuous electron gas
\cite{Moulopoulos92a,Moulopoulos93a}.

\subsection{Application to inorganic $\rho=\frac{1}{2}$ materials}

The thrust of our work has been to understand within a unified theoretical 
approach the variety of exotic insulating states that are proximate to the 
superconducting state in 2:1 cationic or 1:2 anionic 2D CTS. We have shown 
here that the peculiarities of these materials originate from the unique 
behavior of $\rho=\frac{1}{2}$ in the presence of both strong e-e 
interactions and lattice frustration. It is conceivable that the complex 
behavior of apparently unrelated inorganic families can be understood 
within the same broad theoretical approach. We list two such classes of 
materials below to point out the hitherto unnoticed similarities 
between them and $\rho=\frac{1}{2}$ CTS.

\subsubsection{Layered cobaltates}

Layered cobaltates Na$_x$CoO$_2$ have attracted wide attention because of 
their 2D structure, tunable carrier concentration, and the occurrence of 
SC \cite{Takada03b}. The Co ions form a 2D triangular lattice, are in their low-spin state,
and their valence ranges from Co$^{3+}$ at $x=1$ to Co$^{4+}$ at $x=0$. 
Charge carriers are $S=\frac{1}{2}$ holes on the Co$^{4+}$ sites and the hole 
density $\rho=1-x$. Trigonal distortion
splits the occupied $t_{2g}$ orbitals into $e_g^\prime$ and $a_{1g}$ orbitals. LDA
calculations have suggested that although for large $x$ the $e_g^\prime$ orbitals occur below the
Fermi level and an $a_{1g}$-only description is valid, this description breaks down at small $x$
where $e_g^\prime$ orbitals can be nearly degenerate \cite{Singh00a,Lee05b}. In contrast, 
correlated-electron calculations find that the $a_{1g}$ -- $e_g^\prime$ energy separation is 
positive and relatively
large for all $x$ \cite{Bourgeois09a,Landron08a}, suggesting that low energy excitations
can likely be described within $a_{1g}$-only single-band models.

The temperature dependent magnetic susceptibility 
$\chi(T)$ shows a peculiar $\rho$-dependence within the family, with small 
$\rho$ (large $x$) exhibiting strongly correlated behavior and large 
$\rho$ (small $x$) exhibiting weakly correlated behavior \cite{Foo04a}. 
We have recently shown that $\rho$-dependent $\chi(T)$, exactly as seen in the 
cobaltates, is expected within the single-band extended Hubbard model on a triangular lattice 
\cite{Li10b}. Equally interestingly, $\chi(T)$ behavior in Na$_x$CoO$_2$ is very 
similar to that in the family of CTS as a whole, where also $\chi(T)$ 
shows a systematic $\rho$-dependence that is understood within the 
single-band extended Hubbard model \cite{Mazumdar83a}. It is tempting to 
compare now the superconducting states in hydrated cobaltates and CTS with 
this apparent similarity in mind.

Although superconductivity in Na$_x$CoO$_2$ $\cdot$ $y$H$_2$O occurs at $x 
\simeq 0.35$ \cite{Takada03b}, it is now established that the Co-ion 
valency here is determined not only by the Na content, but also by 
H$_3$O$^+$ ions. There have been several reports that SC here occurs over 
a very narrow range of hole density $\rho$, and that maximum $T_c$ occurs 
at or very close to Co-ion valency 3.5+, corresponding to 
$\rho=\frac{1}{2}$ \cite{Barnes05a,Sakurai06a,Banobre-Lopez09a}. If this is confirmed from future 
experimental work, it would appear that like the CTS, cobaltates are yet 
another example of a frustrated 2D $\rho=\frac{1}{2}$ superconductor, 
suggesting that the mechanism of SC in the two families is related.

\subsubsection{$\rho=\frac{1}{2}$ spinels}

The B sublattice in spinel compounds AB$_2$X$_4$ form a frustrated
three-dimensional (3D) pyrochlore lattice and usually consist of
transition metal cations that possess partially filled $t_{2g}$
$d$-orbitals. For integer occupancies of $d$-electrons per $B$ cation
the geometrical degeneracy of the underlying lattice is often lifted
by orbital ordering (OO), leading to formation of spin-singlet
dimers. Only four of the many spinel compounds are superconducting, of
which three have effective carrier density $\rho=\frac{1}{2}$:
LiTi$_2$O$_4$, CuRh$_2$S$_4$, and CuRh$_2$Se$_4$. In LiTi$_2$O$_4$
there is one $d$-electron per two Ti$^{3.5+}$ ions; in CuRh$_2$S$_4$
and CuRh$_2$Se$_4$ the Rh$^{3.5+}$ ions have average $d$-hole
occupancy of $\frac{1}{2}$.  Jahn-Teller instability or OO can lead
to occupancy of the same $t_{2g}$ orbitals, making the filled bands
exactly $\frac{1}{4}$-filled, as in the superconducting organic CTS
(the apparently different spinel superconductor CuV$_2$S$_4$, also has
noninteger number of $d$-electrons per V$^{3.5+}$ ion.)

Yet another similarity between the organics and the $\rho=\frac{1}{2}$ 
spinels is the proximity of the superconducting state to exotic 
semiconducting states. Thus CuIr$_2$S$_4$ and LiRh$_2$O$_4$, both 
isoelectronic with CuRh$_2$S$_4$, undergo MI transitions that are 
accompanied by CO. In CuIr$_2$S$_4$ the Ir ions are charge-ordered as 
Ir$^{3+}$-Ir$^{3+}$-Ir$^{4+}$-Ir$^{4+}$ along specific directions 
\cite{Radaelli02a}. This would correspond exactly to the 
$\cdots$0011$\cdots$ CO in our notation here. As in the 2D PEC, these 
inorganic 3D systems are spin-gapped due to the formation of 
Ir$^{4+}$-Ir$^{4+}$ singlet bonds. It is interesting to recall that 
similar singlet bonds between Ti$^{3+}$ ions had been proposed \cite{Chakraverty78a,Lakkis76a} 
many years back within bipolaron theories of Ti$_4$O$_7$ and 
LiTi$_2$O$_4$. As explicitly shown in our work here, the singlet bond 
formation is a natural consequence of frustration and 
$\frac{1}{4}$-filling. Work is currently in progress to extend the PEC 
concept to the checkerboard lattice, thought to be 2D equivalents of the 
pyrochlore lattice, multiple orbitals per site \cite{Clay10a}.

\section{Conclusions}

In summary, there is an extraordinarily strong tendency to form
spin-singlets in systems with charge carrier concentration precisely
$\frac{1}{2}$.  Naturally, in $\rho=\frac{1}{2}$ this spin-singlet
state is accompanied by CO. The stability of the PEC derives from the
commensurability of the PEC at $\rho=\frac{1}{2}$.  In the anisotropic
triangular lattice, the PEC consists of charge arrangements
$\cdots$1100$\cdots$ in two directions and $\cdots$1010$\cdots$ in the
third direction.  Thus although $\rho=\frac{1}{2}$ in principle is
incommensurate on the triangular lattice, ``separately commensurate''
periodic charge arrangements are nevertheless possible.

\section{Acknowledgments}

We thank M. Abdel-Jawad, H. Fukuyama, K. Kanoda, and R. Kato for 
helpful discussions.  This work was supported by the US Department of
Energy grant DE-FG02-06ER46315. RTC thanks the University of Arizona
for support while on sabbatical.


\begin{thebibliography}{111}
\bibitem[{fru()}]{frustration-note}
\bibinfo{note}{Note that the ratio \protect{$t^\prime/t$} commonly used to
  measure the degree of frustration in an effective \protect{$\rho=1$} model is
  not the same as the \protect{$t^\prime/t$} considered here for the
  \protect{$\rho=\frac{1}{2}$} lattice.}
\end{thebibliography}
\end{document}